%% file: main.tex
\PassOptionsToPackage{table}{xcolor}
\documentclass[sigconf]{acmart}

\input{_frontmatter}

\definecolor{mGreen}{rgb}{0,0.6,0}
\definecolor{mGray}{rgb}{0.5,0.5,0.5}
\definecolor{mPurple}{rgb}{0.58,0,0.82}
\definecolor{backgroundColour}{rgb}{0.95,0.95,0.92}

\newcommand\YAMLkeystyle{\color{magenta}\ttfamily\scriptsize}
\newcommand\YAMLcolonstyle{\color{mGreen}\mdseries}
\newcommand\YAMLvaluestyle{\color{mPurple}\mdseries}

\lstdefinestyle{YAMLStyle}{
    backgroundcolor=\color{backgroundColour},
    basicstyle=\YAMLkeystyle,                                 % assuming a key comes first
    sensitive=false,
    morestring=[b]',
    morestring=[b]",
    stringstyle=\YAMLvaluestyle\ttfamily,
    moredelim=[l][\color{orange}]{\&},
    moredelim=[l][\color{magenta}]{*},
    moredelim=**[il][\YAMLcolonstyle{:}\YAMLvaluestyle]{:},   % switch to value style at :
    captionpos=b,
    keepspaces=true,
    title=\lstname,
    showspaces=false,
    showstringspaces=false,
    showtabs=false,
    tabsize=2
}

\lstdefinestyle{PythonStyle}{
    backgroundcolor=\color{backgroundColour},
    commentstyle=\color{mGreen}\bf,
    keywordstyle=\color{magenta},
    numberstyle=\tiny\color{mGray},
    stringstyle=\color{mPurple},
    basicstyle=\ttfamily\scriptsize,
    breakatwhitespace=false,
    breaklines=true,
    captionpos=b,
    keepspaces=true,
    title=\lstname,
    showspaces=false,
    showstringspaces=false,
    showtabs=false,
    tabsize=2,
    language=Python
}

\begin{document}

%% Title
\title{Secure API-Driven Research Automation to Accelerate Scientific Discovery}

%% Authors
\input{_authors.tex}

\renewcommand{\shortauthors}{Skluzacek et al.}

%% Abstract
\begin{abstract}
The Secure Scientific Service Mesh (S3M) provides API-driven infrastructure to accelerate scientific discovery through automated research workflows. By integrating near real-time streaming capabilities, intelligent workflow orchestration, and fine-grained authorization within a service mesh architecture, S3M revolutionizes programmatic access to high performance computing (HPC) while maintaining uncompromising security. This framework allows intelligent agents and experimental facilities to dynamically provision resources and execute complex workflows, accelerating experimental lifecycles, and unlocking the full potential of AI-augmented autonomous science. S3M signals a new era in scientific computing infrastructure that eliminates traditional barriers between researchers, computational resources, and experimental facilities.
\end{abstract}



%% Keywords

\maketitle

%% Sections
\input{sec_introduction}

\input{sec_s3m}
\input{sec_sdk}
\input{sec_conclusion}

{
\medskip\noindent \small
\textbf{Acknowledgments.} This research used resources of the Oak Ridge Leadership Computing Facility at the Oak Ridge National Laboratory, supported by the Office of Science of the U.S. Department of Energy under Contract No. DE-AC05-00OR22725. The authors acknowledge the guidance and expertise of Verónica G. Melesse Vergara in these efforts. 
}

% \begin{acks}
% To Robert, for the bagels and explaining CMYK and color spaces.
% \end{acks}

%%
%% The next two lines define the bibliography style to be used, and
%% the bibliography file.
\bibliographystyle{ACM-Reference-Format}
\bibliography{references}

\end{document}

%% file: _frontmatter.tex
\usepackage{colortbl}
\usepackage{listings}
\usepackage{graphicx}
\usepackage{makecell}
\usepackage{multirow}
\usepackage{caption}

\definecolor{lightgray}{gray}{0.95}
\definecolor{codegray}{gray}{0.95}
\definecolor{codeblue}{rgb}{0.26, 0.3, 0.6}
\definecolor{codegreen}{rgb}{0,0.5,0}
\definecolor{codered}{rgb}{0.6,0,0}

\definecolor{yamlkey}{rgb}{0,0.5,0} % dark green

\definecolor{keywordblue}{rgb}{0.13,0.13,1}
\definecolor{commentgreen}{rgb}{0,0.5,0}
\definecolor{stringred}{rgb}{0.6,0.1,0.1}

\lstdefinestyle{codeblock}{
    backgroundcolor=\color{gray!20},
    basicstyle=\scriptsize\ttfamily,
    numbers=left,
    numberstyle=\tiny,
    numbersep=5pt,
    frame=single,
    rulecolor=\color{black},
    xleftmargin=2mm,
    framexleftmargin=3mm,
    framexrightmargin=3mm,
    breaklines=true,
    showstringspaces=false
}

\lstdefinestyle{mystyle}{
    backgroundcolor=\color{codegray},
    commentstyle=\color{codegreen},
    keywordstyle=\color{codeblue}\bfseries,
    numberstyle=\tiny\color{gray},
    stringstyle=\color{codered},
    basicstyle=\ttfamily\footnotesize,
    breaklines=true,
    keepspaces=true,
    numbers=left,
    numbersep=5pt,
    showstringspaces=false,
    tabsize=2,
}

\lstdefinelanguage{yaml}{
  morekeywords={[1]kind,spec,templates,tasks,name,templateRef,template,dependencies,arguments,parameters,value},
  keywordstyle={[1]\color{yamlkey}},
  morecomment=[l]{\#},
  morestring=[b]",
  morestring=[b]',
  sensitive=false
}

\acmConference[PEARC '25]{Practice and Experience in Advanced Research Computing}{July 2025}{Columbus, OH, USA}
\acmYear{2025}
\copyrightyear{2025}
\acmISBN{979-8-4007-1398-9/25/07}
\acmDOI{10.1145/3708035.3736072}

%% file: _authors.tex
% Authors not yet ordered by contribution
\settopmatter{authorsperrow=4}

\author{Tyler J. Skluzacek}
\affiliation{%
  \institution{Oak Ridge National Lab}
  \city{Oak Ridge}
  \state{TN}
  \country{USA}
}
\email{skluzacektj@ornl.gov}

\author{Paul Bryant}
\affiliation{%
  \institution{Oak Ridge National Lab}
  \city{Oak Ridge}
  \state{TN}
  \country{USA}
}
\email{bryantpj@ornl.gov}

\author{A.J. Ruckman}
\affiliation{%
  \institution{Oak Ridge National Lab}
  \city{Oak Ridge}
  \state{TN}
  \country{USA}
}
\email{ruckmanaj@ornl.gov}

\author{Daniel Rosendo}
\affiliation{%
  \institution{Oak Ridge National Lab}
  \city{Oak Ridge}
  \state{TN}
  \country{USA}
}
\email{rosendod@ornl.gov}

\author{Suzanne Prentice}
\affiliation{%
  \institution{Oak Ridge National Lab}
  \city{Oak Ridge}
  \state{TN}
  \country{USA}
}
\email{prenticest@ornl.gov}

\author{Michael J. Brim}
\affiliation{%
  \institution{Oak Ridge National Lab}
  \city{Oak Ridge}
  \state{TN}
  \country{USA}
}
\email{brimmj@ornl.gov}

\author{Ryan Adamson}
\affiliation{%
  \institution{Oak Ridge National Lab}
  \city{Oak Ridge}
  \state{TN}
  \country{USA}
}
\email{adamsonrm@ornl.gov}

\author{Sarp Oral}
\affiliation{%
  \institution{Oak Ridge National Lab}
  \city{Oak Ridge}
  \state{TN}
  \country{USA}
}
\email{oralhs@ornl.gov}

\author{Mallikarjun Shankar}
\affiliation{%
  \institution{Oak Ridge National Lab}
  \city{Oak Ridge}
  \state{TN}
  \country{USA}
}
\email{shankarm@ornl.gov}

\author{Rafael Ferreira da Silva}
\affiliation{%
  \institution{Oak Ridge National Lab}
  \city{Oak Ridge}
  \state{TN}
  \country{USA}
}
\email{silvarf@ornl.gov}

\thanks{Notice: This manuscript has been authored by UT-Battelle, LLC, under contract DE-AC05-00OR22725 with the US Department of Energy (DOE). The publisher acknowledges the US government license to provide public access under the DOE Public Access Plan (\url{http://energy.gov/downloads/doe-public-access-plan}).}

%% file: sec_introduction.tex
\section{Introduction}

Modern scientific research increasingly demands seamless integration between experimental and computational facilities, their computing resources, and data management systems to enable autonomous discovery~\cite{skluzacek2024towards}. The emerging paradigm of ``self-driving autonomous laboratories" requires programmatic research interfaces that can coordinate complex workflows that span multiple facilities without (or with minimum) human intervention~\cite{aww2024}. Researchers have traditionally relied on manual methods---logging into compute clusters via SSH, submitting batch jobs, and asynchronously retrieving data, but these approaches fundamentally limit the potential for near-real-time experiment steering, active data analysis, and on-demand resource allocation; functionalities needed for next-generation science.

The rise of generative AI models and reinforcement learning agents capable of scientific reasoning~\cite{hu2024review,cappelloauroragpt} has created new imperatives for the experimental infrastructure. To create truly autonomous AI-based experimentation, researchers must programmatically integrate their research capital---instruments, compute resources, and data repositories---with their AI training, testing, and inference pipelines~\cite{subramanian2024closing, chard2023globus}. 
Implementing such interfaces poses not only technical challenges but also major security and policy concerns~\cite{etz2025interactive}. Although HPC facilities employ strict authentication frameworks to protect resources and data, these same protections create barriers for automated systems, laboratory instruments, edge devices, and AI agents that need to trigger computations in response to experimental results or predictive insights.

The absence of standardized, secure mechanisms to orchestrate workflows between experimental and computational facilities results in fragmented solutions, communication inefficiencies, and missed opportunities for AI-accelerated scientific discovery. The Oak Ridge Leadership Computing Facility's (OLCF) Secure Scientific Service Mesh (S3M) addresses these challenges by providing a facility API---the first of its kind to leverage a flexible service mesh architecture---that enables authenticated external systems and intelligent agents to securely provision resources, stream data, and trigger compute jobs dynamically. This architecture ensures modularity, scalability, and policy-driven security enforcement across computational services. In this paper, we present our work-in-progress architecture, API components, security model, and user interfaces of S3M, demonstrating how this infrastructure enables a new generation of autonomous scientific workflows at OLCF.

% \medskip
\noindent\textbf{\emph{Concise Perspective on Related Scientific APIs.}}
S3M extends previous work on scientific APIs. The Superfacility API~\cite{enders2020cross} provides RESTful interfaces to HPC resources, allowing experiments to transfer data to compute facilities and trigger analysis jobs. FirecREST~\cite{cruz2020firecrest} offers a RESTful web API infrastructure that connects scientific gateways to HPC systems. Globus Flows~\cite{chard2023globus} provides automation of the research process through the cloud-hosted execution of flows on heterogeneous resources. Tapis~\cite{stubbs2021tapis} is a platform for distributed computational research that offers fine-grained authorization, data management, and code execution capabilities. SCEAPI~\cite{rongqiang2017sceapi} provides a unified RESTful API for accessing HPC resources in multiple Chinese supercomputing centers, supporting authentication, file transfer, and job management. S3M distinguishes itself through a service mesh architecture, allowing highly customizable services, fine-grained policy enforcement, and dynamic routing, capabilities not possible in traditional API gateways. It introduces advanced streaming for low-latency data exchange enabling near real-time decision making, seamless workflow orchestration, and support for custom API extensions. These features are invaluable for high-security experimental and computational facilities that need to quickly process and act on incoming data streams.

%% file: sec_s3m.tex
\section{The Scientific Service Mesh for Automated Science}
\label{sec:s3m}

\begin{figure}[t]
    \centering
    \includegraphics[width=\linewidth]{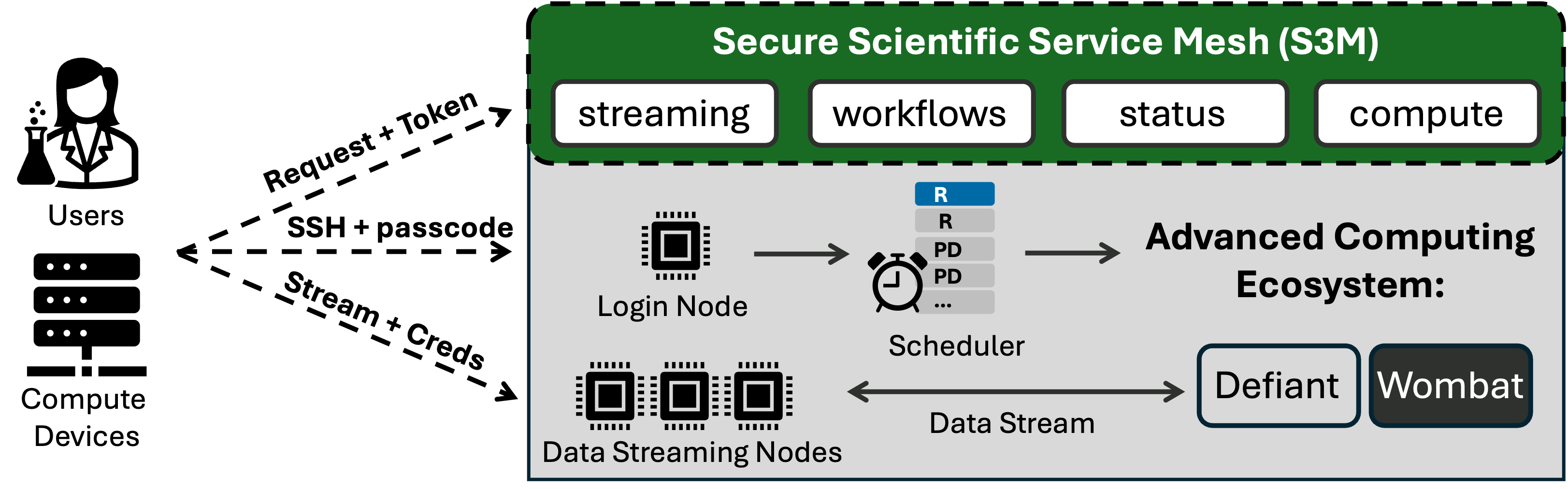}
    \caption{S3M Architecture diagram. Automated and human clients can make requests to S3M if they have an S3M Access Token. Authorized users have access to the various APIs hosted in front of OLCF resources. S3M then handles the provisioning of streaming nodes and communication with the compute resources' schedulers. }
    \label{fig:architecture}
\end{figure}

The Secure Scientific Service Mesh (S3M) provides programmatic access to OLCF's HPC resources by integrating distinct services, managed policies, and fine-grained authorization into a unified framework. Built on a flexible service mesh architecture using OpenShift, deployed in OLCF’s Slate clusters~\cite{OLCF_Slate}, S3M enables scientific instruments, workflows, and intelligent agents to interact securely with HPC systems while maintaining strong security through layered policy-as-code access controls and project-scoped authentication. A \textit{service mesh} is an infrastructure layer that facilitates secure and efficient communication between services in a distributed system, abstracting networking complexity while enforcing authentication, authorization, and traffic management policies~\cite{service_mesh}. In S3M, this architecture ensures modularity, scalability, and security, allowing independent management of core services, including advanced streaming for near real-time data exchange, workflow orchestration across facility boundaries, status monitoring, and compute scheduling. This design, visualized in Fig.~\ref{fig:architecture}, allows new capabilities to be explored without disrupting existing OLCF infrastructure, which is particularly valuable for AI-driven experimental feedback loops that must operate within the OLCF's HPC environment.

% tried to make less redundant with prior paragraph. 
S3M relies on Istio~\cite{istio}, an open-source service mesh platform that provides fine-grained traffic management, security, and observability features, to enforce multilayered validation through authentication, authorization, and policy compliance checks. The traceability features in Istio offer a comprehensive view of all requests and behavior within the mesh, supporting compliance and security auditing. Users obtain project-scoped authentication tokens with strictly defined permissions, and S3M validates every request against project allocations and resource policies before execution. This approach enables dynamic resource provisioning and workflow automation, while preserving the integrity of the OLCF's computing environment.

\subsection{Secure API Communications Framework}

S3M provides an extensive set of APIs, accessible at both gRPC\allowbreak+\allowbreak Protobuf~\cite{grpc, protobuf}
 and RESTful JSON endpoints, that enable researchers to interact with OLCF resources programmatically. Each endpoint serves a specific purpose within the scientific workflow automation ecosystem, from monitoring resource availability to submitting large and complex compute tasks. The core API components, summarized in Table~\ref{tab:endpoints}, are designed to support diverse scientific needs while maintaining strict access controls and tightly integrating a wide range of access policies. In the following, we describe in more detail two of S3M's unique APIs: Streaming and Workflows.

\begin{table*}[!ht]
    %\scriptsize
    \centering
    \begin{tabular}{p{2cm} p{11cm}}
        \toprule
        \textbf{API Endpoint} & \textbf{Description} \\
        \midrule
        /status & Provides resource availability information, including overall system status, specific resource states, and scheduled downtimes. \\
        
        \rowcolor{gray!30}
        /compute & Supports job submission and management, allowing users to submit, track, and cancel compute jobs on available resources. \\
        
        /streaming & Manages data streaming resources, enabling provisioning, listing, and deallocation of Redis or RabbitMQ instances for low-latency scientific workflows. \\
        
        \rowcolor{gray!30}
        /environment & Retrieves dependency and runtime environment information for computing workflows. \\
        
        /tokens & Manages secure API access tokens for authenticated service interactions. \\
        
        \rowcolor{gray!30}
        /workflows & Facilitates workflow automation by allowing submission, status retrieval, and cancellation of complex workflows across heterogeneous computing resources. \\
        \bottomrule
    \end{tabular}
    \caption{Core S3M API Endpoints and Their Functionality.}
    \label{tab:endpoints}
    % \vspace{-20pt}
\end{table*}

The \emph{\textbf{Streaming API}} is one of S3M’s foundational capabilities, enabling real-time science applications by allowing instruments, agents, and workflows to dynamically stream data to and from HPC systems. While many facility APIs provide well-defined interfaces for accessing individual resources, they typically lack capabilities to connect compute jobs with experiment control applications. As a result, researchers must manually deploy and maintain external messaging services to facilitate data exchange. The Streaming API enables researchers to provision RabbitMQ or Redis messaging services on dedicated high-throughput streaming nodes near computational resources through simple API calls. This feature is crucial for interactive science applications that require low-latency data exchange between experimental facilities and compute resources, enabling near real-time decision-making and instrument feedback loops. By abstracting the complexity of message broker management behind a unified interface, the Streaming API simplifies the development of these data-intensive scientific workflows. 

\begin{figure}[!ht]
    \centering
    \includegraphics[width=\linewidth]{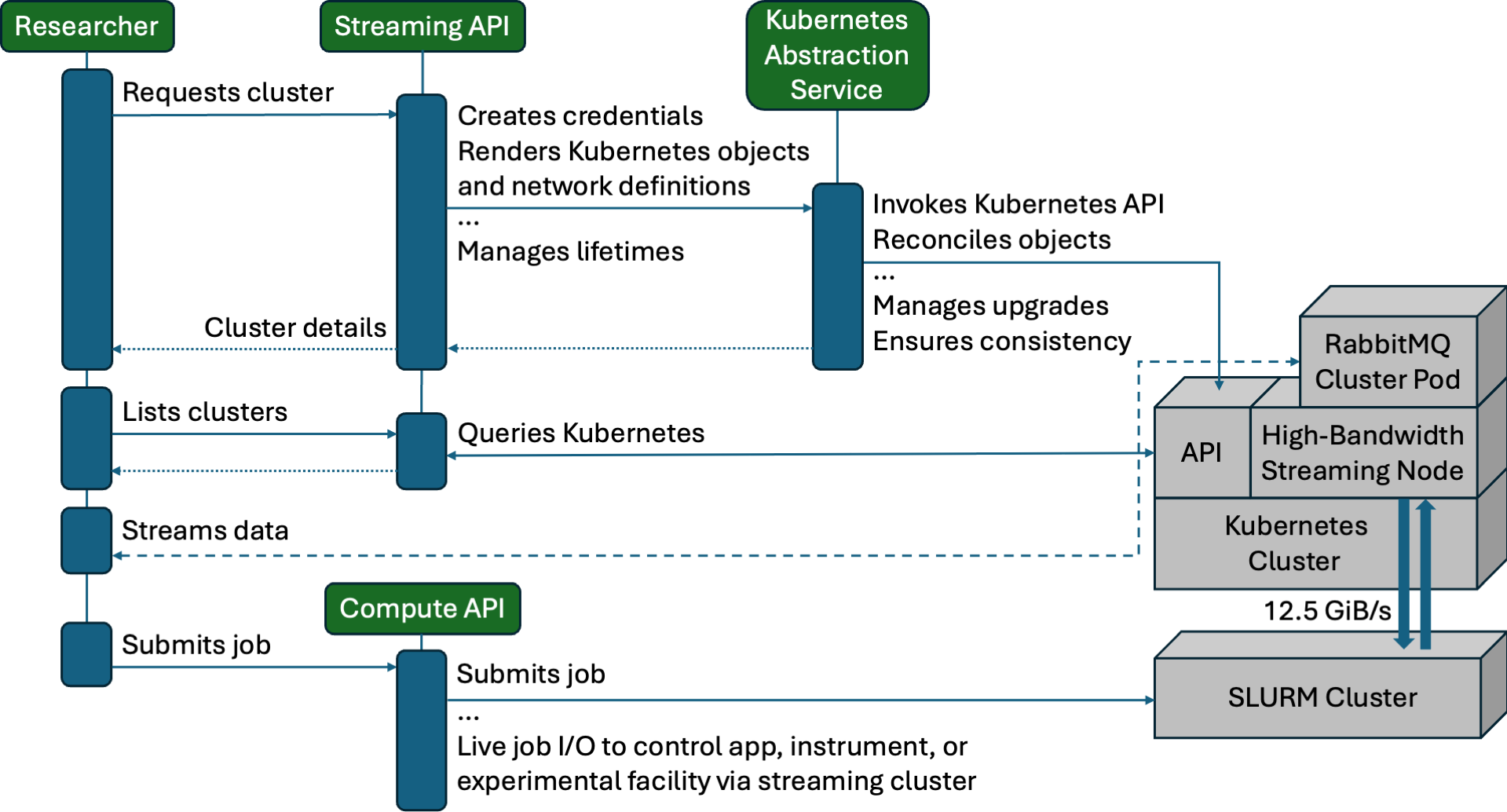}
    \caption{S3M Streaming Service Interaction Flow. This diagram shows how researchers interact with S3M services to provision streaming infrastructure and interface with a compute cluster. The Stream Manager creates Kubernetes objects while the Abstraction Service deploys the RabbitMQ cluster. Researchers can then develop data pipelines with these resources.}
    %\vspace{-5px}
    \label{fig:streaming_compute}
\end{figure}

Previously, OLCF users needing live interactions with compute jobs faced a tedious, multistep process: (1)~securing an allocation on one of our Kubernetes application clusters or finding their own hardware to host a message broker; (2)~installing and configuring their broker; (3)~requesting firewall exceptions to enable communication between instruments and facilities; and (4)~maintaining the health and security of their broker. Furthermore, externally hosted brokers were often physically distant from control applications or compute resources, leading to increased latency and inconsistent throughput. The Streaming API eliminates these challenges by automating broker provisioning in secure, OLCF-approved environments that are proximal to both OpenShift application clusters and computational resources. Fig.~\ref{fig:streaming_compute} illustrates the interaction flow between the researcher, the S3M Streaming and Compute APIs, and the underlying infrastructure that enables these capabilities.

The \emph{\textbf{Workflow API}} extends S3M's capabilities by integrating with Argo Workflows~\cite{argo-workflows}, enabling researchers to orchestrate complex, multistep scientific processes with minimal manual intervention. By supporting the submission of Argo Workflow Templates, this API allows users to define, reuse, and share sophisticated execution pipelines that seamlessly incorporate custom endpoints and data management tools while handling parallel workflow invocations; data dependencies and artifact management; and fault tolerance. This abstraction layer significantly reduces the developer error and execution overhead, allowing scientists to focus on research objectives rather than managing computational logistics. The Workflow API represents a critical component for autonomous science, where reproducible and efficient processing chains across distributed resources are essential to discovery.

% *** Tyler edit line ***

\subsection{Layered Authentication and Authorization Framework}

S3M enforces a multi-tiered security architecture to validate all remote client interactions with OLCF resources. Each API request must include an authorization token generated by a valid user through our trusted web portal. This ensures that only users with appropriate project access and a sufficient account status can generate credentials. All network traffic is encrypted at the gateway, mitigating risks from credential interception or adversary-in-the-middle attacks.
%To create the token, 
%users perform a strong identity verification process that requires authentication appropriate to each enclave: RSA key-based authentication via a hardware crypto card or mobile device for OLCF's more secure \textit{moderate} enclave, and password authentication for OLCF's \textit{open} enclave. 

Once a request reaches S3M, the system validates the user's identity, project affiliations, and resource access against the OLCF infrastructure. Requests that fail these checks are immediately rejected, preventing unauthorized access before reaching internal services. If a request passes the authentication and authorization layers, it is routed to the relevant service, such as compute job submissions to Slurm schedulers~\cite{schedmd}. To manage security across internal communications, all S3M services communicate over mutual TLS (mTLS), reducing unauthorized access risks within the system. Extensive logging captures details for all requests in addition to internal communication throughout the service mesh, ensuring full traceability for compliance and anomaly detection. This model minimizes implicit trust and strengthens access controls, supporting broader efforts to adopt zero-trust principles across OLCF infrastructure.

%Responses to clients follow the same secure pathway, with all interactions logged for auditability and security oversight. These logs capture authentication attempts, authorization decisions, and API activity, ensuring full traceability for compliance and anomaly detection---both essential in a national laboratory computing environment.

% removed, per Paul's request
% As future work, S3M is prototyping policy-driven access control for fine-grained authorization decisions. This would allow context-aware enforcement of time-based restrictions, resource-specific permissions, and adaptive rate limits, with policies dynamically updated to meet evolving requirements.

% S3M enforces a multi-tiered security architecture to validate all remote client interactions with OLCF resources. Each API request must include an authorization token generated by a valid OLCF user on our trusted web portal. This ensures that only verified OLCF users with appropraite project access can generate credentials. These tokens are subject to comprehensive authrozation procedues  All network traffic is encrypted at the gateway, mitigating risks from credential interception or adversary-in-the-middle attacks. 

%% file: sec_sdk.tex
\section{Programmatic Research Interface}

The Software Development Kit (SDK) provides a simple Python interface for both human researchers and automated systems to interact with OLCF. By encapsulating complex API interactions into intuitive service classes, the SDK eliminates common implementation challenges around authentication, request formatting, or error handling, allowing scientists to focus on research objectives rather than infrastructure mechanics. The package is installed via pip and securely manages the authentication token using an environment variable to prevent credential exposure.

\begin{figure}[ht]
    \centering
    \lstset{style=PythonStyle, language=Python}
    \begin{lstlisting}[caption={Provisioning and managing a data streaming cluster using the S3M Python SDK.}, label={listing:stream_job}]
from olcf_s3m_api.client import OLCFAPIClient
from olcf_s3m_api.streaming import StreamingService

client = OLCFAPIClient(token=os.environ['S3M_TOKEN'])
service = StreamingService(service_name="rabbitmq", 
                           api_client=client)

status = service.start_cluster(
    cluster_name = "my-rmq-cluster",
    node_count   = 1,
    cpu_count    = 4,
    ram_gib      = 4
)

# calls to RabbitMQ cluster using Pika library
. . . 
service.stop_cluster(cluster_name="man-cluster")
    \end{lstlisting}
\end{figure}

The streaming service example in Listing~\ref{listing:stream_job} illustrates the dynamic provisioning of a dedicated messaging infrastructure for the exchange of near real-time data between scientific instruments and computational resources. With just a few lines of code, researchers can orchestrate the entire messaging infrastructure lifecycle: from dynamically provisioning a fully configured RabbitMQ cluster with precise CPU and memory allocations, to seamlessly transmitting experimental data through established channels, to automatically decommissioning resources upon completion. This capability is particularly valuable for automated scientific workflows that require low-latency communication for experimental steering and adaptive decision-making based on emerging computational results. Listing~\ref{listing:argo} illustrates the description of a multitask directed acyclic graph (DAG) workflow in Argo. This workflow reuses predefined templates (e.g., Listing~\ref{listing:stream_job}) to deploy the streaming service, submit the compute job, and check the job status. Such workflow automation and template reusability help to lower the barrier for reproducing complex experiments.
\begin{figure}[ht]
    \centering
    \lstset{style=YAMLStyle, language=Yaml}
    \begin{lstlisting}[caption={Argo Workflow template to automatically orchestrate streaming service deployment, job submission, and status monitoring via the S3M /workflows API endpoint.}, label={listing:argo}]
kind: Workflow
spec:
  templates:
    dag:
      tasks:
        - name: deploy-streaming-service
          templateRef:
            template: deploy-streaming
        - name: submit-job
          dependencies: [deploy-streaming-service]
          templateRef:
            template: submit-job
        - name: check-job-status
          dependencies: [submit-job]
          templateRef:
            template: check-job-status
          arguments:
            parameters:
              - name: JOB_ID
                value: "{{tasks.submit-job.outputs.params.JOB_ID}}"
    \end{lstlisting}
\end{figure}

%% file: sec_conclusion.tex
\section{Future Directions for Scientific Automation}

S3M advances autonomous API-driven scientific workflows by using a service mesh architecture to enable secure and scalable interactions between researchers, AI agents, and HPC resources. By unifying fine-grained authorization, dynamic resource provisioning, and low-latency data streaming under a cohesive framework, S3M establishes the foundation for next-generation scientific automation. Although currently available to internal users on select clusters, our roadmap includes
refining authentication policies with input from diverse science projects, integrating advanced workflow management systems, publishing comprehensive SDK documentation, and expanding to external user access. As S3M evolves toward deployment on the OLCF's exascale Frontier supercomputer, this framework will dramatically accelerate experimental lifecycles, enable adaptive research workflows in near real-time, and ultimately transform how AI-augmented science operates by eliminating traditional barriers between instruments, computational resources, and researchers.